\documentclass[aps,pre,twocolumn,superscriptaddress]{revtex4-1}
\usepackage{amsmath}
\usepackage{amsfonts}
\usepackage{amssymb}
\usepackage{epsfig}
\usepackage{graphicx}
\usepackage{color}
\renewcommand{\phi}{\varphi}
\newcommand{\be}{\begin{equation}}
\newcommand{\ee}{\end{equation}}
\newcommand{\ba}{\begin{align}}
\newcommand{\ea}{\end{align}}
\newcommand{\red}[1]{\textcolor{black}{#1}}

\newcommand{\ppdif}[2]{\frac{{\partial}^2 #1}{\partial^2 {#2}}}

\newcommand{\ave}[1]{\langle {#1} \rangle}

\begin{document}

\title{Large-scale structure of randomly jammed \red{spheres}}

\author{Atsushi Ikeda}
\affiliation{Graduate School of Arts and Sciences, 
University of Tokyo, Komaba, Tokyo 153-8902, Japan}

\author{Ludovic Berthier}
\affiliation{Laboratoire Charles Coulomb, UMR 5221, CNRS and Universit\'e
de Montpellier, 34095 Montpellier, France}

\author{Giorgio Parisi}
\affiliation{Dipartimento di Fisica,
Universit\`a degli studi di Roma La Sapienza, 
Nanotec-CNR, UOS Rome,
INFN-Sezione di Roma 1, Piazzale A. Moro 2, 00185, Rome, Italy}

\date{\today}

\begin{abstract}
We numerically analyse the density field of three-dimensional randomly jammed packings of 
\red{monodisperse soft frictionless} spherical particles, paying special attention to
fluctuations occurring at large lengthscales.
We study in detail the two-point static structure factor
at low wavevectors in Fourier space. We also analyse the nature of the density
field in real space by studying the large-distance behavior
of the two-point pair correlation function, of density 
fluctuations in subsystems of increasing sizes, and of the 
direct correlation function. We show that such real space analysis 
can be greatly improved by introducing
a coarse-grained density field to disentangle genuine large-scale 
correlations from purely local effects.
Our results confirm that both Fourier and real space signatures 
of vanishing density fluctuations at large scale are absent, indicating
that randomly jammed packings are not hyperuniform. 
In addition, we establish that the pair correlation function displays 
a surprisingly complex structure at large distances, 
which is however 
\red{not compatible with the long-range negative correlation of 
hyperuniform systems but} fully compatible 
with an analytic form for the structure factor. This implies
that the direct correlation function is short-ranged,
as we also demonstrate directly. 
Our results reveal that density fluctuations 
in jammed packings do not follow the behavior expected 
for random hyperuniform materials, but display instead a more
complex behavior.
\end{abstract}


\maketitle
\section{Introduction}

There is a growing research effort to understand 
the structure of hyperuniform materials, for which 
density fluctuations at large lengthscales display 
unusual 
properties~\cite{torquato2003,donev2005,zachary2009,jiao2014,more_sal}. 
This problem raises fundamental
questions about structural properties of complex
and non-equilibrium systems~\cite{hexner,jack2015,tjhung2015,denis}, 
and has potential applications
to devise novel materials with non-conventional 
properties~\cite{mat1,mat2,mat3,mat4,mat5}.

In particular, it was conjectured 
that randomly jammed packings of spherical particles 
are hyperuniform~\cite{torquato2003}. 
More precisely, the static structure factor $S(k)$ 
of jammed  assemblies of hard spheres has been 
fitted in the low-$k$ regime to a non-analytic functional 
form~\cite{donev2005,hopkins2012} 
\begin{equation}
S(\vec{k}) = A + B | \vec{k} |,
\label{eq:claim}
\end{equation}
where $A$ and $B$ are numerical constants. Because $A$ is
numerically found to be very small, the hyperuniform property
that $S( \vec{k} \to 0) = 0$ follows. 
The hyperuniformity of jammed packings 
was first observed in monodisperse system~\cite{donev2005,silbert2009}, 
then also in binary mixtures~\cite{zachary2011,berthier2011}, 
and even in polydisperse 
systems~\cite{berthier2011,kurita2010,kurita2011}. 
However, recent simulations have challenged 
this finding~\cite{wu2015,ikeda2015,atkinson2016b}, 
and provided evidence that randomly jammed packings 
are not strictly hyperuniform, in the sense that  $S( \vec{k} \to 0) > 0$.

In addition to the vanishing wavevector 
limit, the non-analytic linear-$k$ behavior in 
Eq.~(\ref{eq:claim}) has several direct consequences
at large but finite lengthscale. First, it implies 
that the pair correlation function $g(r)$ converges to 
its asymptotic limit from negative values
and as a power law~\cite{donev2005}, $g(r) - 1 \propto - r^{-4}$,
a behavior which has never been observed directly, to our knowledge. 
A second consequence is that density fluctuations 
display anomalous behavior at large scale, 
so that the variance of the fluctuations of the number density
in a subsystem of size $R$ scales (in three dimensions) as 
$\langle \Delta D(R) \rangle \sim R^{-4}$, instead of the 
weaker $R^{-3}$ scaling expected from the central limit 
theorem~\cite{torquato2003,zachary2009}. Deviations 
from $R^{-3}$ behavior were recently reported 
in experimental work~\cite{dreyfus2015}.
A third consequence is that the direct correlation 
function $c(r)$ should become long-ranged and 
exhibit a power law decay \cite{torquato2003}, $c(r) \sim r^{-2}$. Numerical 
evidence in favour of this behavior was recently published~\cite{atkinson2016}. 
 
To conclude that randomly jammed sphere packings 
are disordered hyperuniform materials, one should ideally establish that 
Eq.~(\ref{eq:claim}) together with all its direct consequences 
are consistently observed in a single system.  
In particular, the power law decay of the pair correlation
function is difficult to observe, because it is masked by 
an additional oscillatory 
behavior with a wavelength corresponding approximately to 
the particle diameter $\sigma$, 
arising from purely local correlations. These oscillations tend to obscure the 
power law decay at large distances~\cite{donev2005}. 
In the same vein, it was  
noted that these local correlations may also affect the scaling 
of the density fluctuations $\langle \Delta D(R) \rangle$
to produce an apparent anomalous scaling that could be unrelated to 
hyperuniformity~\cite{wu2015}.

In this work we simultaneously measure $S(k)$, $g(r)$,
$c(r)$, and $\langle \Delta D(R) \rangle$
for random packings of spherical particles. We use 
large-scale simulations in three dimensions to 
determine the nature of density fluctuations at large distances.
In addition, we introduce 
a coarse-grained density field to successfully 
disentangle genuine large-scale correlations of the density 
field from more local effects, and this allows us to 
accurately determine the large-scale behavior of $g(r)$
and $\langle \Delta D(R) \rangle$. Our results reveal 
a surprisingly complex pattern of density fluctuations at large-scale 
in jammed packings, but the results differ 
from the hyperuniform behavior in Eq.~(\ref{eq:claim})
on two grounds. We find that $S( \vec{k} \to 0)$ does not vanish and 
that the non-analytic linear wavevector dependence does 
not consistently account 
for our large body of numerical results. 

The outline of the paper is as follows. We explain our protocol
to produce jammed packings in Sec.~\ref{sec:preparation}. 
We then study respectively the static structure factor, 
the pair correlation function, the local fluctuations of the 
density and the direct correlation functions 
in Secs.~\ref{sec:sk}, \ref{sec:gr}, \ref{sec:density}, 
and \ref{sec:cr}.
We conclude in Sec.~\ref{sec:conclusion}.

\section{Preparation of jammed packings}
\label{sec:preparation}

We focus on the same model as in our previous 
studies~\cite{ikeda2015,ikeda2013}.  
We consider a system of monodisperse spheres in three dimensions, 
where spheres interact through a pairwise harmonic potential~\cite{durian1995} 
$v(r_{ij}) = \frac{\epsilon}{2} (1-r_{ij}/\sigma) \Theta(\sigma - r_{ij})$ 
where $\Theta (x)$ is the Heaviside step function, $r_{ij}$ is 
the distance between particles $i$ and $j$ 
and $\sigma$ is the particle diameter.
Throughout this work, we use $\sigma$ and $\epsilon$ as the 
units of length and energy. 

We generate configurations of randomly packed spheres by the following 
protocol. We first prepare a random configuration of $N$ spheres in 
a cubic box of dimension $L$ at $\phi=0.8$, 
where $\phi = \frac{\pi N \sigma^3}{6 L^3}$ is the packing fraction.  
The number of particles is $N=512000$ unless otherwise noted
(see Appendix \ref{app:finitesize} for a discussion of finite size 
effects).
Then we use the FIRE algorithm to minimize the potential energy and to 
find the mechanical equilibrium state~\cite{bitzek2006}.
\red{The algorithm is efficient enough that large system sizes 
can be studied.}
Next we gradually decrease the packing fraction in small steps
and minimize the energy after each step to obtain 
the mechanical equilibrium state at the desired packing fraction. 
\red{Our strategy is to find energy minima for configurations 
above jamming, and to observe how the structure of these packings changes 
as density is varied. This approach differs from, and is numerically
much simpler than, studies focusing on packings produced 
precisely at jamming, for which very precise algorithms need to be 
devised~\cite{atkinson2016b,eric}.} 

We repeat these calculations from independent 
initial random configurations and obtain a large number of random packings. 
The number of independent configurations is 240 for the lowest density 
and 960 for the highest density addressed. We carefully tested the convergence
against the number of samples for each measured quantity,
as detailed in Appendix~\ref{app:finitesample}.

We report various correlation functions which are obtained by 
averaging over results obtained for each independently produced  
configuration at a given density. 
Note that these different configurations are characterized by 
distinct values of the critical density for the jamming transition,
thus the distance to the jamming density fluctuates from one sample
to another~\cite{ohern2003}. 
However because our system is sufficiently large, 
the fluctuations of these distances are negligible 
for all the parameters studied in this work, 
thus the sample average at each density can be taken safely. 
Indeed, the standard deviation of the pressure, which gives an 
estimate of the fluctuations of these distances  
is less than 1\% of the average pressure, even at 
the lowest density addressed. 

\section{Static structure factor}
\label{sec:sk}

Let us start by focusing on the static structure 
factor in Fourier space. 
The number density fields in real and reciprocal spaces are 
respectively defined as~\cite{hansen} 
\be
\rho(\vec{x}) = \sum_{i=1}^N \delta(\vec{x} - \vec{R}_i), \ \ \ 
\rho(\vec{k}) = \sum_{i=1}^N e^{i\vec{k} \cdot \vec{R}_i},
\ee
where $\vec{R}_i$ is the coordinate of particle $i$.
Using these fields, we define the density-density correlation functions
\be
G(\vec{r}) = \frac{1}{\rho^2} \ave{\rho(\vec{x}) \rho(\vec{x}+\vec{r})}, 
\ \ \  S(\vec{k}) = \frac{1}{N} \rho(\vec{k}) \rho (-\vec{k}),  \label{eq:gofr}
\ee
where $\ave{ \cdots } = \frac{1}{L^3}\int d^3\vec{x}$ 
is the translational average
and $\rho = \frac{N}{L^3}$ is the number density. 
We denote the spherical averages of 
these correlation functions as $G(r)$ and $S(k)$. 
As in the case of the radial distribution 
function and the static structure factors in liquid states~\cite{hansen}, 
$G(r)$ and $S(k)$ are related through a Fourier transform.

Jammed systems are often said to be hyperuniform~\cite{donev2005}. 
Hyperuniformity is mathematically defined as~\cite{more_sal}
\be
\lim_{k \to 0}S(k) = 0. 
\ee 
Because the static structure factor in this limit is the standard 
deviation of the number of particles, 
this property means that density fluctuations decay 
with increasing the lengthscale of observation
more rapidly than the prediction obtained from the central limit theorem.  
Thus, hyperuniformity implies the existence of 
some sort of long-range correlations in the density field. 
The static structure factor of jammed packings generated by 
fast compressions of hard spheres was studied numerically, 
and is reported to behave as in Eq.~(\ref{eq:claim})
with $A = 6.1 \times 10^{-4}$ and $B = 3.4 \times 10^{-3}$~\cite{donev2005}. 
The observation that $A$ is very small has led to the 
claim that jammed packings are hyperuniform.  
It is also noted that the linear wavevector dependence 
in Eq.~(\ref{eq:claim}) means that $S(k)$ becomes a 
non-analytic function at $k=0$, a behavior which is 
never observed in ordinary liquid states at thermal 
equilibrium~\cite{hansen}.

\begin{figure}
\psfig{file=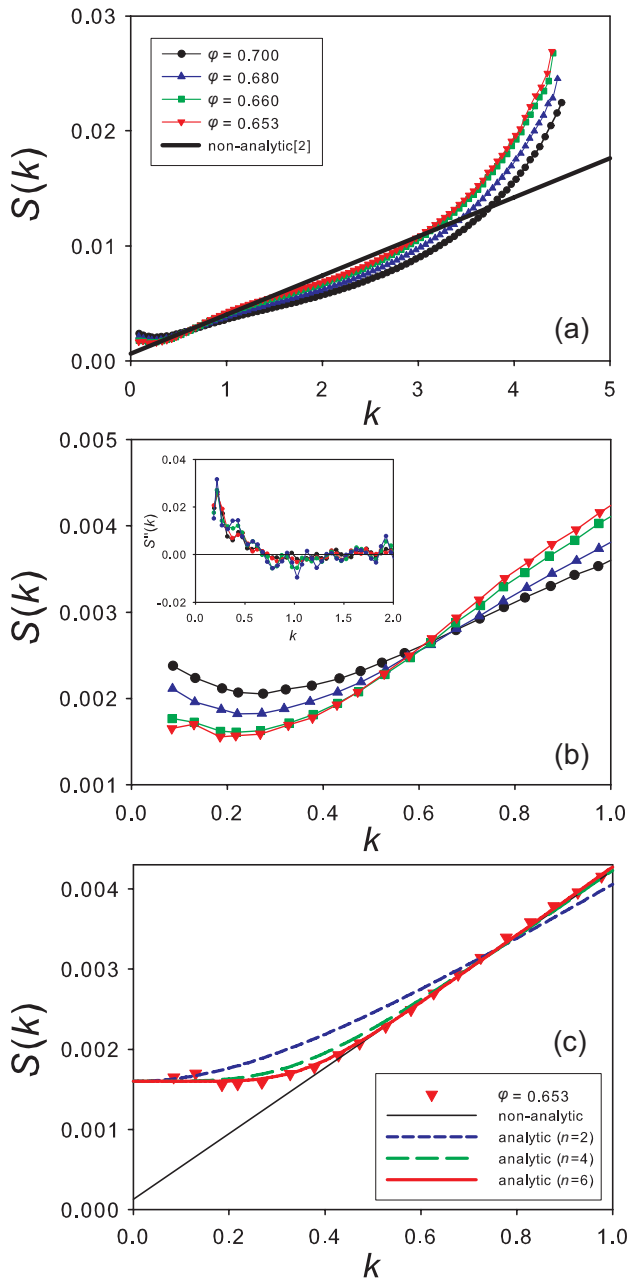,width=8.2cm,clip}
\caption{\red{(a)} Static structure factor at 
$\phi =$ 0.7, 0.68, 0.66 and 0.653 with the non-analytic linear model 
with the parameters optimized in Ref.~\cite{donev2005}. 
\red{(b)} A zoom on the low wavevector behavior of the top panel. 
The inset shows the second derivative of the static 
structure factor with a line highlighting the apparent linear
behavior of $S(k)$. 
\red{(c)} The structure factor at $\phi=0.653$ fitted 
to various functional forms.}
\label{fig:sk}
\end{figure}

We first compute the static structure factor $S(k)$ in the following way. 
We compute the density field $\rho(\vec{k})$ on the lattice points
of the form 
$\vec{k} = (n_1, n_2, n_3)\Delta k$, where the $n_i$'s 
are integers and $\Delta k = \frac{2\pi}{L}$, 
and calculate $S(\vec{k}) = \frac{1}{N} \rho(\vec{k}) \rho(-\vec{k})$ 
on each lattice point. 
\red{Then, we average $S(\vec{k})$ in spherical shells in the reciprocal space with the width $\delta k$ to obtain $S(k)$. 
We used $\delta k = 0.05$ for $N=512000$ and $\delta k = 0.03$ for $N = 4048000$.}

The computed $S(k)$ at $\phi =$ 0.7, 0.68, 0.66 and 0.653 are 
plotted in Fig.~\ref{fig:sk}. 
The pressure values at 
these state points are 0.029, 0.017, 0.0059, and 0.0026, 
thus these states are almost equally separated from each other 
in a logarithmic scale for the 
pressure~\cite{ohern2003}. Irrespective of the density, 
$S(k)$ displays nearly linear behavior in the wide range of wave 
numbers at $k \leq 3$, as previously reported~\cite{donev2005,berthier2011}. 
However, deviations from this linear behavior take place at $k \approx 0.3$
and $S(k)$ becomes nearly constant or even increases slightly 
with decreasing $k$ even further. Such deviations were recently
noted~\cite{wu2015,ikeda2015,atkinson2016b} and they are more clearly
demonstrated by measuring the second derivative $S''(k)$. As
shown in the inset of Fig.~\ref{fig:sk}, $S''(k) = 0$ only holds
for $k > 0.5$ but clear and robust deviations appear at lower 
wavevectors. Our findings show that these deviations 
are not due to the numerical difficulty to produce 
packings precisely at the jamming transition~\cite{atkinson2016b}.

We have confirmed that the existence of this upturn 
at $k \approx 0.3$ is numerically robust 
by changing both the system size and the number of independent configurations
used to perform the ensemble average, as shown in Appendices 
\ref{app:finitesize} and \ref{app:finitesample}.
With decreasing the density towards jamming, 
$S(k)$ slightly increases at large $k$ and slightly decreases at low $k$, 
but this dependence saturates at $\varphi = 0.66$, which is far above jamming.  
This indicates that the jamming criticality, which affects 
many physical quantities, does not dramatically affect 
the low-$k$ behavior of $S(k)$. This conclusion 
was reached in previous work \cite{wu2015,ikeda2015}, and 
originates from the fact 
that the jamming criticality is sensitive to detailed 
features of particle interactions near contact, whereas 
we focus here on large-scale density fluctuations. 

The numerical results for $S(k)$ can not be fitted using 
the non-analytic functional form $S(k) = A + B|\vec{k}|$ at $k < 0.3$. 
In the bottom panel of Fig.~\ref{fig:sk}, we display 
this function with $A = 1.3 \times 10^{-4}$ and $B = 4.1 \times 10^{-3}$
and the $S(k)$ at the lowest density, 
where the fitting parameters $A$ and $B$ are determined 
by a least squared fitting over the interval $k \in [0.4,1.0]$.
 
Instead, the numerical results can be well fitted into the 
empirical analytic function
\be
S(k) = S_0 \left[ 1 + \left( \frac{k \xi}{\pi}\right)^n \right]^{1/n}.
\label{eq:analytic}
\ee
This function converges to the constant value $S \approx S_0$ when 
$k \xi \ll 1$, and 
behaves as a linear function $S \approx S_0 \xi k/\pi$ when $k\xi \gg 1$
and it is thus well-suited to describe the saturation observed in the data
near $k \approx 0.3$.
We display these functions for the values $n=2$, 4 and 6
where $S_0$ is fixed to the value $S_0=0.016$ 
and the remaining fitting parameters are $\xi = 7.32$, 8.26 and 8.38 for 
$n=2$, 4 and 6 respectively, 
are again determined by a least square fitting. 
The convergence of $S(k)$ to the constant  at 
$k \to 0$ is well captured by both functions 
and the sharpness of the upturn at $k \approx 0.3$ is better captured by 
the values $n=4$ and 6. 
Note that the parameter $\xi$ gives an estimate of the wavelength 
corresponding to the wavenumber where the upturn occurs. 

In summary, $S(k)$ shows strong  deviations from the proposed 
non-analytic linear behavior at low-$k$, 
and exhibits a sharp crossover which is well
captured by an empirical analytic function given by Eq.~(\ref{eq:analytic}).

\section{Pair correlation function in real space}
\label{sec:gr}

If $S(k)$ has the non-analytic form $S(k) = A + B|\vec{k}|$, 
then the real space density correlation function $G(r)$ has the interesting 
property that its asymptotic decay becomes~\cite{donev2005} 
\begin{eqnarray}
G(r) \to 1 - \frac{B}{\rho \pi^2 r^4}, \ \ \  r \to \infty.
\end{eqnarray}
Namely, there appears a long-range ``negative'' density correlation 
which does not have any characteristic lengthscale
and is instead algebraic. 
Thus it is interesting to study the large-$r$ behavior of $G(r)$ 
to confirm whether such a correlation exists
in jammed packings, because it represents the direct 
counterpart to the linear $k$ dependence of the static structure factor. 
However numerical analysis of this putative 
power law behavior is very difficult 
because $G(r)$ also has a strong oscillatory behavior which is  
caused by the short-range correlations of the density field.
Physically, this is because the well-separated 
wavevector regimes in the Fourier domain get entangled by the inverse Fourier 
transform. In order to overcome this problem, 
we introduce a coarse-grained density field and study its correlation function.

\subsection{Coarse-grained density field}

The coarse-grained density field $\psi(\vec{x})$ is defined as 
\begin{eqnarray}
\psi(\vec{x}) = \sum_i f(\vec{x} - \vec{R}_i), \ \ \  
f(\vec{x}) = \Bigl( \frac{\delta}{\pi} \Bigr)^{3/2} e^{- \delta |\vec{x}|^2},
\label{eq:filt}
\end{eqnarray}
where $f(\vec{x})$ is a Gaussian function which acts as a 
low-frequency filter, 
and $1/\sqrt{\delta}$ is a lengthscale controlling the filter width. 
Then, the coarse-grained density-density correlation function is defined as 
\begin{eqnarray}
Q(\vec{r}) = \frac{1}{\rho^2} \ave{\psi(\vec{x}) \psi(\vec{x}+\vec{r})},  
\label{eq:qofr}
\end{eqnarray}
and so $Q(\vec{r})$ can be seen as a coarse-grained version of $G(\vec{r})$,
and both functions should become equivalent 
when $\delta \to \infty$. To look into the large-$r$ behavior, 
$Q(\vec{r})$ is more suitable than $G(\vec{r})$, 
because the Gaussian filter in Eq.~(\ref{eq:filt}) should eliminate the 
oscillatory behavior due to short-range correlations, 
if $\delta$ is well-chosen. This statement becomes 
evident when we consider the relation between $Q(r)$ and $S(k)$
\begin{eqnarray}
Q(\vec{r}) = \frac{1}{\rho (2\pi)^3} \int d^3\vec{k} 
\ e^{-i\vec{k} \cdot \vec{r}} f(\vec{k})^2 S(\vec{k}), \label{stoq}
\end{eqnarray}
where $f(\vec{k}) = e^{- k^2/(4\delta)}$ is the Fourier transform of $f(\vec{x})$. 
When $\delta$ is small, $f(\vec{k})$ becomes extremely small at large-$k$, 
thus the integral in Eq.~(\ref{stoq})
is not influenced by the first diffraction peak of 
$S(\vec{k})$ which provides the oscillations in real space. In other words, 
local correlations are filtered out by the Gaussian filter
which leaves the low-$k$ behavior intact.
We use $\delta = 0.3$ (the filter width is $1/\sqrt{\delta} 
\approx 1.83$) unless otherwise noted, 
which gives $f(\vec{k})^2 \approx 10^{-36}$ for $|\vec{k}|= 2\pi$, 
where the first peak of $S(k)$ is located.  
\red{Note that the use of the Gaussian filter Eq.~(\ref{eq:filt}) is similar 
in spirit to the use of ``definition I'' in Ref.~\cite{wu2015} although the Gaussian filter is presumably more efficient to suppress the effect of short-range correlations.}
Because $Q(r)$ converges to 1 in the large-$r$ limit, 
it is convenient to define 
\be
P(\vec{r}) = Q(\vec{r}) -1,
\ee
which corresponds to the coarse-grained version of the total 
correlation function $H(\vec{r})$ used in liquid state theory~\cite{hansen}. 
Hereafter, we denote $P(r)$ the spherically averaged $P(\vec{r})$. 
Note that because $f(\vec{x})$ is properly normalized,
the integration of $P(\vec{r})$ over the whole space gives $S(0)/\rho$.  

Before looking into the numerical results for jammed packings, 
we consider two simple examples of $P(r)$. 
The first is the case when the static structure factor is 
constant $S(k) = S(0)$. 
This simplification can apply for simple liquids 
because $S(k)$ converges rapidly to the constant (compressibility) 
at low-$k$~\cite{hansen}.  
In this case, we can compute the integral in Eq.~(\ref{stoq}) which gives
\begin{eqnarray}
P(r) = \frac{S(0)}{\rho} \Bigl( \frac{\delta}{2\pi} \Bigr)^{3/2} 
e^{- \delta r^2/2}.
\label{eq:pr2} 
\end{eqnarray}
Namely, $P(r)$ is characterized by a single Gaussian 
peak at $r=0$ with a width given by $1/\sqrt{\delta}$. 
The oscillations in $G(r)$ are perfectly suppressed in $P(r)$
by the Gaussian filter, as expected. 
From Eq.~(\ref{eq:pr2}), we observe that the 
integration of $P(r)$ up to $r=1/\sqrt{\delta}$ 
is enough to estimate $S(0)$, showing that  
density fluctuations at the local scale are sufficient to 
recover the macroscopic limit in simple liquids.

\begin{figure}
\begin{center}
\psfig{file=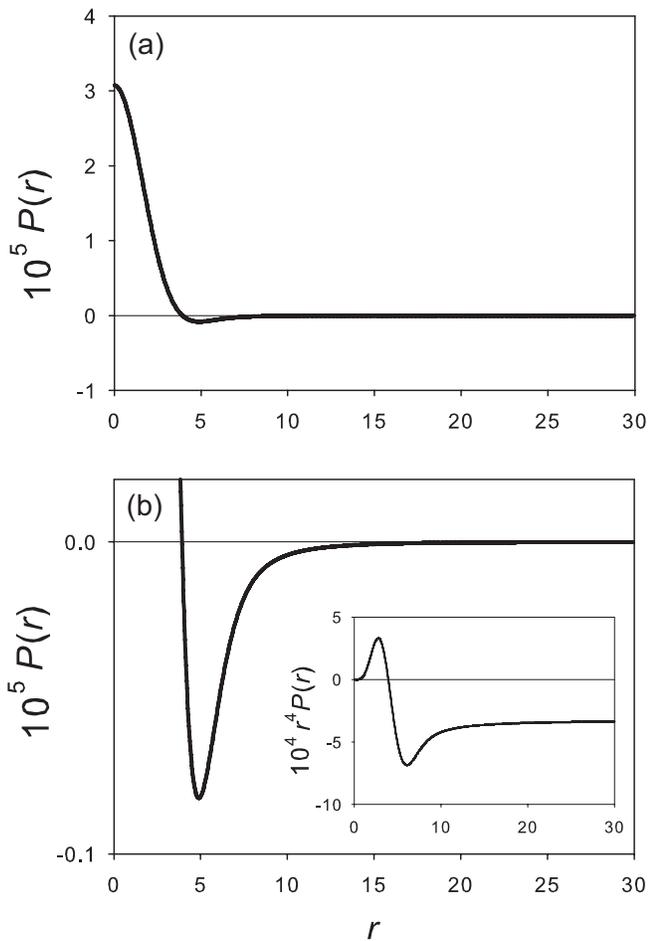,width=8.5cm,clip}
\caption{\red{(a)} Behavior of the coarse-grained density 
correlation function $P(r)$ calculated for a 
non-analytic structure factor of the form 
$S(k) = A + B|\vec{k}|$, with $A$ and $B$ taken from the fit
shown in Fig.~\ref{fig:sk}.
\red{(b)} A zoom on the negative correlation near $r \approx 5$ \red{of (a)}. 
The inset illustrates the $r^{-4}$ power law decay of $P(r)$ 
from negative values as $r \to \infty$.}
\label{fig:pr}
\end{center}
\end{figure}
 
The second example is the non-analytic function $S(k) = A + B|\vec{k}|$,
as in Eq.~(\ref{eq:claim}). 
We insert this function into Eq.~(\ref{stoq}) and evaluate 
the integral numerically, using $A$ and $B$ values 
from the fit shown in Fig.~\ref{fig:sk}.
The obtained $P(r)$ is plotted in Fig.~\ref{fig:pr}. We observe that 
$P(r)$ not only has the positive peak at $r=0$, it also has  
a negative peak near $r \approx 5$. 
The bottom panel of Fig.~\ref{fig:pr} shows that 
the asymptotic decay of $P(r)$ for $r \to \infty$ 
is characterized by the power-law 
\begin{eqnarray}
P(r) \to - \frac{B}{\rho \pi^2 r^4}, \ \ \  r \to \infty,  
\label{eq:asymptotic}
\end{eqnarray}
as expected also for $G(r)$. Thus, in contrast to our 
first example, $P(r)$ is now long-ranged. 
To recover the density fluctuations in the macroscopic limit, 
we must now integrate $P(r)$ up to infinity 
to take into account the negative power law correlation at 
longer lengthscales.  

\subsection{Numerical results}

We now move to the numerical measurement of $P(r)$ in our jammed 
packings. To this end, we evaluate $P(r)$ in the following way. 
We discretize the simulation box into a set of 
lattice points $\vec{x} = (n_1,n_2,n_3) \Delta x$, 
compute the density field $\psi(\vec{x})$ on each lattice point, 
and then calculate the correlation function 
$P(\vec{r}) = \frac{1}{\rho^2} \ave{\psi(\vec{x}) \psi(\vec{x}+\vec{r})} - 1$, 
followed by translational and spherical averages 
to finally get $P(r)$. It should be noted that 
$P(r)$ does not converge to 0 as $r \to \infty$
in a finite system with a fixed number of particles 
(the same remark applies to $G(r)$ in equilibrium 
systems~\cite{salacuse1996}), and 
thus special care should be exercised  about finite size effects,
as explained in Appendix \ref{app:finitesize}. 

\begin{figure}
\begin{center}
\psfig{file=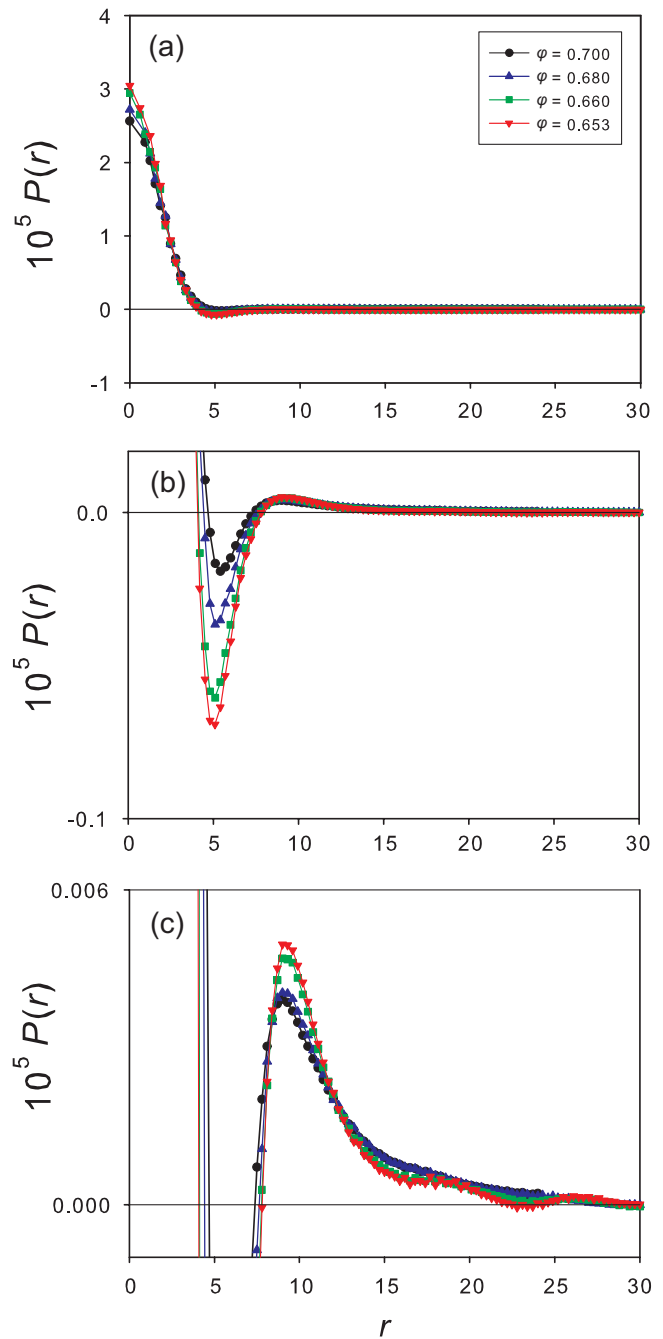,width=8.5cm,clip}
\caption{The measured coarse-grained density 
correlation function $P(r)$ at $\varphi = 0.7$, 0.68, 0.66, 0.653. 
using three different \red{representations} to emphasize
the peak at $r = 0$ \red{(a)}, the negative dip near $r \approx 5$ \red{(b)},
and the positive correlation near $r \approx 10$ \red{(c)}.}
\label{fig:prnum}
\end{center}
\end{figure}

The obtained $P(r)$ is plotted in Fig.~\ref{fig:prnum} for various 
packing fractions. The first important observation from that plot
is that the oscillatory behavior found for $G(r)$ 
with a short wavelength approximately given by $\sigma$,  
is very efficiently suppressed for $P(r)$, which allows us to carefully 
discuss the large-distance behavior of the pair correlation
function in real space. This validates our idea to 
introduce a coarse-grained density field. 

For all densities, $P(r)$ has the peak at $r=0$ which expresses 
the density fluctuations at the microscopic lengthscale. We also 
observe that $P(r)$ has a negative peak at $r \approx 5$. 
This negative peak is very similar to the one of $P(r)$ 
obtained from $S(k) = A + B|\vec{k}|$ (compare with Fig.~\ref{fig:pr}),
which affects density fluctuations in the macroscopic limit. 
However, a new feature emerges at larger distances, since the 
negative peak in the measured $P(r)$ is followed by a
positive correlation peak near $r \approx 10$. 
This behavior is in sharp contrast with the 
non-analytic case $S(k) = A + B|\vec{k}|$, 
where density correlations remain negative with a power law decay.
It is also clear that these basic features of $P(r)$ do not 
depend on the chosen density, and in particular on the distance 
to the critical density of jamming.
Thus, the negative correlation at $r \approx 5$ and 
the appearance of the positive correlation at $r \approx 10$ 
are not associated to the jamming criticality but are  instead
robust features of jammed packings. 

\begin{figure}
\begin{center}
\psfig{file=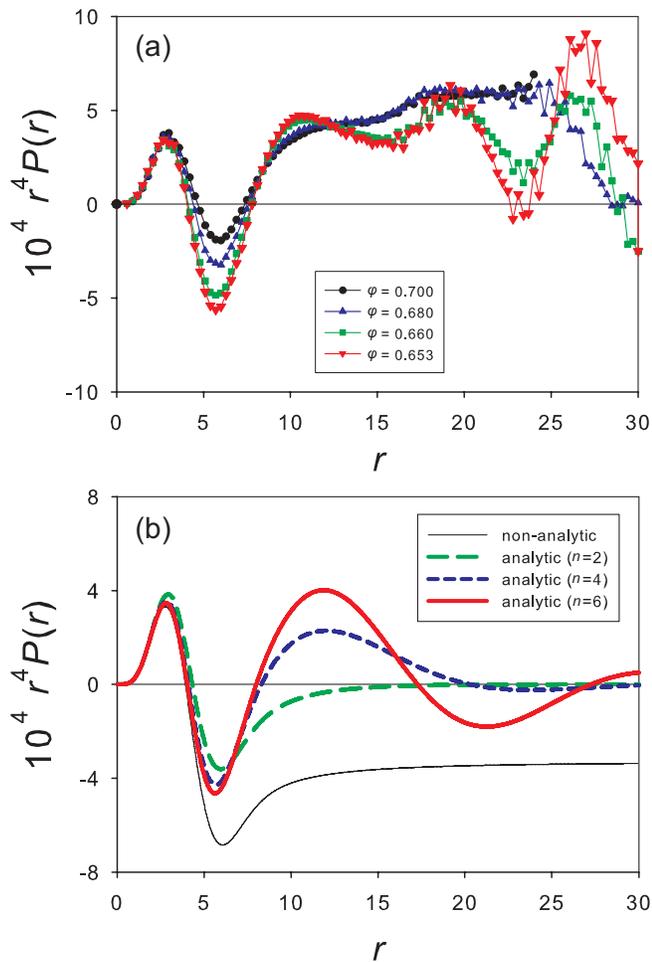,width=8.5cm,clip}
\caption{Behavior of $r^4 P(r)$ for 
\red{(a)} the measured $r^4P(r)$ at $\varphi = 0.7$, 
0.68, 0.66, and 0.653, and  
\red{(b)} the calculated $r^4 P(r)$ 
from non-analytic [Eq.~(\ref{eq:claim})] 
and analytic [Eq.~(\ref{eq:analytic})] models of $S(k)$.
The negative correlation predicted from the non-analytic 
model is not observed in the measured $P(r)$, which 
exhibits a positive correlation which is well captured 
by the analytic models with $n \geq 4$.
}
\label{fig:pr4}
\end{center}
\end{figure}

We look at the large-$r$ behavior of $P(r)$ more closely, and
plot the quantity $r^4 P(r)$ in Fig.~\ref{fig:pr4}. 
As expected from the structure of $P(r)$ in Fig.~\ref{fig:prnum}, we
find that $r^4 P(r)$ is first positive, 
then becomes negative near $r \approx 5$, and 
finally positive again for larger distances.  
In particular, we find that the large distance behavior of 
$P(r)$ is qualitatively distinct from the non-analytic 
behavior expected for hyperuniform materials, 
Eq.~(\ref{eq:asymptotic}), and it is actually even more complicated
than expected.  

To understand the origin of the second positive peak near $r \approx 10$,  
we also plot in Fig.~\ref{fig:pr4} the quantity
$r^4 P(r)$ obtained from analytic models of $S(k)$ 
with various $n$, from Eq.~(\ref{eq:analytic}). 
Although the analytic model with $n=2$ does not show the second positive peak, 
the analytic model with $n \geq 4$ reproduces the peak very well.
Recall that the value of $n$ 
corresponds in Fourier space to the sharpness of the 
crossover behavior near $k \approx 0.3$, 
it is then clear that the second positive peak in $P(r)$ or $r^4 P(r)$ 
is the direct consequence of this low-$k$ feature in $S(k)$. 
Indeed, the second positive peak is located at a 
lengthscale comparable to value of $\xi$ extracted from 
fitting $S(k)$, which corresponds to the wavenumber of the crossover. 
Moreover, the expression for $S(k)$ from the analytic model has 
complex singularities (branch points) at 
$(k \pi)/\xi =(-1)^{1/n}$ and for $n>2$ these singularities 
correspond to an exponentially-decaying 
oscillatory behaviour at large distances.
Note that the position of the second positive peak depends 
on the parameter $\delta$ in the coarse-graining in general, 
but it becomes independent of $\delta$ 
when $\sqrt{\delta}$ is sufficiently large compared to the wavenumber 
of the kink in $S(k)$. We have checked that our 
choice $\delta = 0.3$ satisfies this condition.

We finally note that the second positive peak of $r^4 P(r)$ does not 
decay rapidly with distance and seems to persist even at 
very large-$r$, at least up to $r \approx 20$. 
Because the behavior of $r^4 P(r)$ at $r \to \infty$ is related to a
linear wavevector of $S(k)$ as $k \to 0$, 
a positive value for $r^4 P(r)$ at large-$r$ is related 
to the upturn of $S(k)$ at very low $k$, see the inset 
in Fig.~\ref{fig:sk}. 
To check whether $r^4 P(r)$ finally goes to zero at $r \to \infty$
would require studying an even larger system size, 
which is beyond the scope of the present work. 

\section{Density fluctuations in subsystems}
\label{sec:density}

Density fluctuations in jammed particle systems are often discussed 
in terms of the number density fluctuations within a spherical 
cavity~\cite{torquato2003,zachary2009,dreyfus2015,wu2015}. 
The physical reason is that 
the suppression of density fluctuations expected for
hyperuniform materials at low wavevector
should correspond to an anomalous
behavior of density fluctuations measured in subsystems of increasing sizes 
in real space~\cite{torquato2003,zachary2009}. Since we found above 
that $S(k)$ in jammed packings is incompatible with 
hyperuniform behavior, mathematical consistency requires that 
local density fluctuations should display normal scaling, but this 
expectation contrasts with earlier reports~\cite{donev2005,dreyfus2015}.

The number density in a spherical cavity is defined as
\begin{eqnarray}
D (\vec{x};R) = \frac{3}{4 \pi R^3} \int_R d^3\vec{r} \ 
\rho(\vec{x} + \vec{r}), 
\end{eqnarray}
where $R$ is the radius of a spherical cavity centered at position
$\vec{x}$. 
The variance of the fluctuations of the density is then defined  as
\begin{eqnarray}
\ave{\Delta D(R)} \equiv \ave{D (\vec{x};R)^2} - \ave{D (\vec{x};R)}^2. 
\end{eqnarray}
This variance is directly related to $S(k)$ as follows 
\begin{eqnarray}
\ave{\Delta D(R)} 
&=& \frac{9\rho}{16 \pi^2 R^6} \int \frac{d^3\vec{k}}{(2 \pi)^3} \ S(k) 
\Bigl( \int_R d\vec{r} \ e^{-i \vec{k} \cdot \vec{r}} \Bigr)^2 \nonumber \\
&=& \frac{9\rho}{4 \pi R^3} \int^{\infty}_0 \frac{dk}{k} S(k) J_{3/2} (kR)^2,  
\label{numflu}
\end{eqnarray}
where $J_n(x)$ is the Bessel function of order $n$.
From the central limit theorem, 
we expect that this variance is proportional 
to $R^{-3}$, namely that it  scales as the inverse volume of the cavity
when $R \to \infty$. 
However, in a hyperuniform system, 
it is known that this variance becomes proportional to $R^{-4}$, 
namely that it is dominated by the surface term.  
For jammed packings, this variance was measured 
and was shown to follow the surface term, 
which was considered as an evidence that 
jammed particle systems are hyperuniform~\cite{torquato2003,zachary2009}. 

However, it was pointed out that this quantity 
is considerably affected by the surface term 
due to the slow decay of the Bessel function in Eq.~(\ref{numflu}), 
and that extremely large values of $R$ are required to 
eliminate this effect~\cite{wu2015}. Physically, the reason 
is again that large wavevectors in Eq.~(\ref{numflu}) contribute significantly 
to the final result even though they originate from purely local
correlation effects. To detect hyperuniformity, one
should instead focus on low wavevectors corresponding to 
density fluctuations at large distances. It is therefore pertinent to
the analyse local fluctuations of the coarse-grained density 
$\psi(\vec{x})$, because this quantity
will not be influenced by local correlations.

Following the above definitions for the number density $\rho(\vec{x})$,
we now define the coarse-grained density in a spherical cavity as
\begin{eqnarray}
\Psi (\vec{x};R) = \frac{3}{4 \pi R^3} \int_R d^3\vec{r} \ 
\psi(\vec{x} + \vec{r}), 
\end{eqnarray}
and we define the variance of this density as
\begin{eqnarray}
\ave{\Delta \Psi (R)} \equiv \ave{\Psi  (\vec{x};R)^2} - 
\ave{\Psi  (\vec{x};R)}^2. 
\end{eqnarray}
This variance is also directly related to $S(k)$ as 
\begin{eqnarray}
\ave{\Delta \Psi (R)} = \frac{9\rho}{4 \pi R^3} \int^{\infty}_0 
\frac{dk}{k}  f(k)^2 S(k) J_{3/2} (kR)^2.  \label{psiflu}
\end{eqnarray}
Compared to Eq.~(\ref{numflu}), 
the term $f(k)^2$ now appears in the integral 
over wavevectors and its effect is to reduce the influence 
of large wavevectors and thus of short-range correlations
over the measurement of local fluctuations of the density. 
For a truly hyperuniform materials, both 
$\ave{\Delta D(R)} $ and $\ave{\Delta \Psi (R)}$
would have the same $R^{-4}$ anomalous scaling behavior. 

\begin{figure}
\begin{center}
\psfig{file=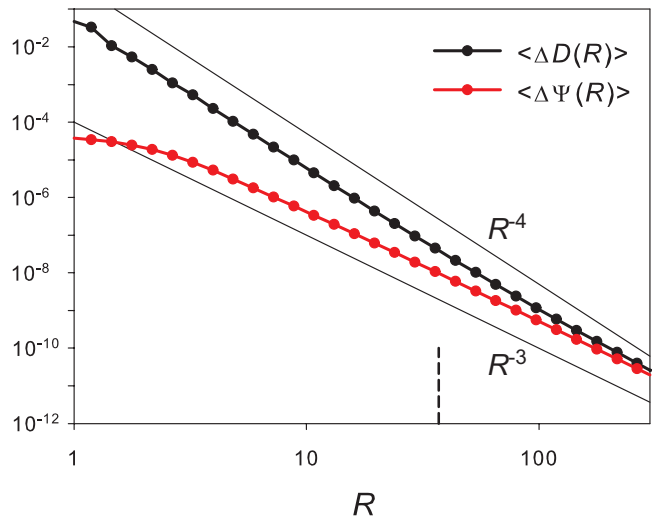,width=8.5cm}
\caption{Measured variance of density fluctuations in a spherical 
cavity of radius $R$ at $\varphi=0.66$. 
$\ave{\Delta D(R)} $ is the variance of the microscopic 
number density $\rho(\vec{x})$, 
and $\ave{\Delta \Psi(R)} $ is the variance 
of the coarse-grained density $\psi(\vec{x})$. 
The apparent anomalous scaling of $\ave{\Delta D(R)}$
originates from short-ranged correlations which are filtered
out in $\ave{\Delta \Psi(R)}$. \red{This quantity 
obeys the scaling expected from
the central limit theorem. This shows that density 
fluctuations are not suppressed at large scale in jammed packings. 
The vertical dashed line indicates 
half of the dimension of the simulation box. }}
\label{fig:delta}
\end{center}
\end{figure}

Plugging the simulation results for $S(k)$ at $\varphi=0.66$ 
into Eq.~(\ref{numflu}) and (\ref{psiflu})
and numerically evaluating the integrals, 
we calculate $\ave{\Delta D (R)}$ and $\ave{\Delta \Psi (R)}$ and plot 
them in Fig.~\ref{fig:delta}. 
\red{In the numerical integrations, we linearly extrapolated the simulation results for $S(k)$ to $k \to 0$. 
Therefore, this calculation gives reliable results for $R \leq L/2$, where $L=74$ is the dimension of the simulation box $(N=512000)$, but the reliability at larger $R$ of course depends on the reliability of this extrapolation.}
The fluctuations of the number density $\ave{\Delta D (R)}$ show 
the (apparent) hyperuniform scaling $R^{-4}$ for a wide range of cavity 
sizes from $R=1$ to 100. 
On the other hand, the fluctuations of the coarse-grained density 
$\ave{\Delta \Psi (R)}$ 
show a normal scaling with inverse volume $R^{-3}$ over almost 
the entire range of cavity sizes. Both functions 
seem to become equivalent at 
very large cavity sizes only, above $R \approx 200$.
 
This result means that the apparent
hyperuniform behavior of $\ave{\Delta D (R)}$ 
mainly results from
short-range correlations of the number density 
appearing in the integral 
(\ref{numflu}), and is thus unrelated to a suppression of 
density fluctuations occurring at large scale. 
The normal $R^{-3}$ scaling found for the coarse-grained density 
fluctuations is consistent with the behavior of the structure factor 
$S(k)$ which does not show signs of hyperuniformity.

We also find that the integral in Eq.~(\ref{numflu}) is strongly 
affected by the large-$k$ region of $S(k)$ 
and a numerical integration up to $k=30$ is required to obtain a 
converged result for the quantity $\ave{\Delta D (R)}$.  
This observation confirms our conclusion and is also consistent with the 
results reported in Ref.~\cite{wu2015}. This means that 
future exploration of hyperuniformity in 
particle systems based on local fluctuations of the density
should employ a coarse-grained density field in order to 
more directly detect large-scale effects and to decrease
the influence of short-range correlations.

\section{Direct correlation function}
\label{sec:cr}

We finally focus on the direct correlation function $C(r)$.
This function is defined through the Ornstein-Zernike equation~\cite{hansen} 
\begin{eqnarray}
H(\vec{r}) = C(\vec{r}) + \rho \int d^3\vec{r}' C(\vec{r} - 
\vec{r}') H (\vec{r}'), 
\end{eqnarray}
where $H(\vec{r}) = G(\vec{r}) - 1$ is the total correlation function. 
The convolution integral contained in this equation
simplifies in the reciprocal space, and this relation 
can be rewritten as 
\begin{eqnarray}
C(k) = \frac{S(k) - 1}{\rho S(k)}, \label{dcf}
\end{eqnarray}
where $C(k)$ is the Fourier transform of $C(r)$. 
We use Eq.~(\ref{dcf}) to convert our numerical data for 
$S(k)$ into $C(k)$, 
and we then calculate $C(r)$ by performing an inverse 
Fourier transform. 
Both $S(k)$ obtained directly from the simulations 
and from the various models of $S(k)$ are used. 
Note that the Fourier transform from $C(k)$ to $C(r)$ 
requires a special care because $C(k)$ has a $k^{-1}$ 
behavior at large $k$ which stems from the divergence of 
the first peak in $G(r)$ near jamming~\cite{ohern2003,ikeda2013a}, 
and is thus a purely local effect again. To control the convergence of
this numerical calculation, 
we multiply $C(k)$ with a Gaussian window function, $w(k) = e^{- (k/k_w)^2}$,
before performing the numerical Fourier transform to ensure its convergence. 
We checked that all values $k_w \geq 5$ give essentially the same 
result for $C(r)$ at distances $r > 1$. Thus, we 
fix $k_w = 100$ in the following. 

\begin{figure}
\begin{center}
\psfig{file=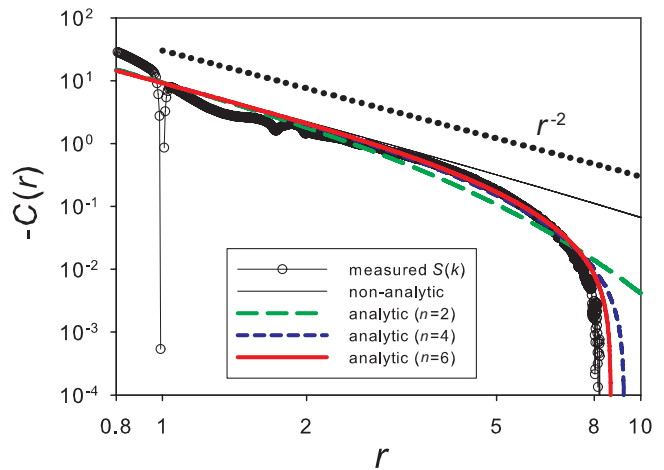,width=8.5cm}
\caption{Direct correlation function $C(r)$
obtained from numerical simulations at $\phi=0.66$
and calculated from various models for $S(k)$ from 
Eqs.~(\ref{eq:claim}, \ref{eq:analytic}). The analytic models 
with $n \geq 4$ reproduce the measured behavior very well.}
\label{fig:cr}
\end{center}
\end{figure}

In Fig.~\ref{fig:cr}, we show $C(r)$ obtained from 
direct simulations and from the various models for $S(k)$. 
\red{This function displays a sharp dip near $r=1$
which reflects the physics at contact, as also seen in 
hard spheres~\cite{atkinson2016}.}  
There are two features at large distances to be observed 
in the direct correlation $C(r)$ obtained from the 
simulations. First, $C(r)$ follows a power law scaling 
$|C(r)| \sim r^{-2}$ over a modest range of distances  $r \approx 1-6$, 
but it then exhibits a sudden drop near $r \approx 8$ where it changes sign.
This power-law scaling was related 
in earlier work~\cite{atkinson2016} to the linear behavior of $S(k)$, 
because $S(k) \sim k$ directly implies 
from Eq.~(\ref{dcf}) that $C(k) \sim k^{-1}$, and thus $C(r) \sim r^{-2}$. 

To understand the observed behavior of $C(r)$ in more 
detail, it is instructive to use our various models for 
$S(k)$ to obtain ``synthetic'' $C(r)$ functions, which can then be compared to 
the simulation results. 
The non-analytic model $S(k) = A + B|\vec{k}|$ indeed gives 
the expected power-law scaling $r^{-2}$ of $C(r)$ 
up to very large distances, and it shows no sign of a sudden drop.
On the other hand, the analytic model with $n\geq 4$ 
almost perfectly reproduces both the power-law scaling at very short distances
$r < 6$ and also the drop near $r \approx 8$. The amplitude
of these functions then goes rapidly to zero at large distance 
and shows no sign of a power law behavior (not shown).
 
These observations imply that the drop of $C(r)$ 
originates again from the saturation of $S(k)$ observed at very low 
wavector, $k \approx 0.3$, in Fig.~\ref{fig:sk}. 
Note in addition that the drop in $C(r)$ takes place at $r \approx 8$, 
which is almost identical to $\xi$, 
the lengthscale corresponding to the wavenumber of crossover in $S(k)$, 
and is also close to the second positive peak in $P(r)$ 
discussed in Fig.~\ref{fig:prnum}.
Our conclusion is then that a hyperuniform material should 
exhibit a $r^{-2}$ power law scaling of the direct correlation 
function up to very large distances, whereas a non-hyperuniform material
with an analytic structure factor would instead display a strong 
drop of $C(r)$ at a finite distance. The second 
behavior is the one that is most consistent with our numerical results. 

Comparing our results to an earlier report in Ref.~\cite{atkinson2016}, 
we conclude that our simulation results for $C(r)$ are 
almost identical to the ones obtained for compressed hard spheres. 
This suggests that $S(k)$ for such hard sphere packings is 
presumably equivalent to our own data, and presumably contains also  
a crossover to a non-hyperuniform behavior at low wavevector.
Thus, our conclusions on the large-scale structure of jammed configurations 
may also apply for the configurations analyzed in Ref.~\cite{atkinson2016}, 
despite the difference in the preparation protocols. A more 
quantitative comparison between the two sets of results would be interesting.

\section{Conclusion}
\label{sec:conclusion}

We have studied the nature of spatial correlations
of the density field at large distances in random packings of spheres. 

Our results show that a mathematical description of the 
density-density correlations 
in terms of a random hyperuniform structure does not 
describe jammed packings very well. Instead, 
density fluctuations do not vanish asymptotically at large distances, 
and they do not display an anomalous lengthscale-dependence 
when measured in subsystems. The real space counterparts 
of the proposed non-analytic linear wavevector dependence
of the structure factor are not consistent with our 
numerical observations. We conclude therefore that density 
fluctuations display a complex behavior at large scale, 
but this behaviour appears analytic and not hyperuniform. 

It would be interesting to understand the complex structure 
of such random packings from analytic theory, for instance
by extending previous work~\cite{parisi2010,charbonneau2014} 
to also describe large-scale 
structural properties. Another interesting question is whether
our findings affect the relevance 
of random sphere packings from the viewpoint of material 
science~\cite{mat1,mat2,mat3,mat4,mat5}, or whether 
such details do not affect the transport 
properties of such materials. In that case, the structure 
of jammed packings would still be of considerable theoretical
interest, as it forms a puzzle related to the 
structure of amorphous solids at large lengthscales. \red{This 
question has been rarely} addressed in the field of glassy materials,
where only structural features from 
local to medium range are more typically discussed.

\acknowledgments
The research leading to these results has received funding from the 
European Research Council under the European Unions
Seventh Framework Programme (FP7/2007-2013)/ERC
Grant Agreement No. 306845. This work was supported
by a grant from the Simons Foundation 
(\# 454933, Ludovic Berthier,
 \# 454949, Giorgio Parisi)
\red{and the JSPS KAKENHI (\# 16H04034, Atsushi Ikeda).}
The calculations have been done in 
Supercomputing Division, Information Technology Center, 
The University of Tokyo and
Research Institute for Information Technology, 
Kyushu University.

\appendix

\section{Finite-size effects}
\label{app:finitesize}

\begin{figure}
\begin{center}
\psfig{file=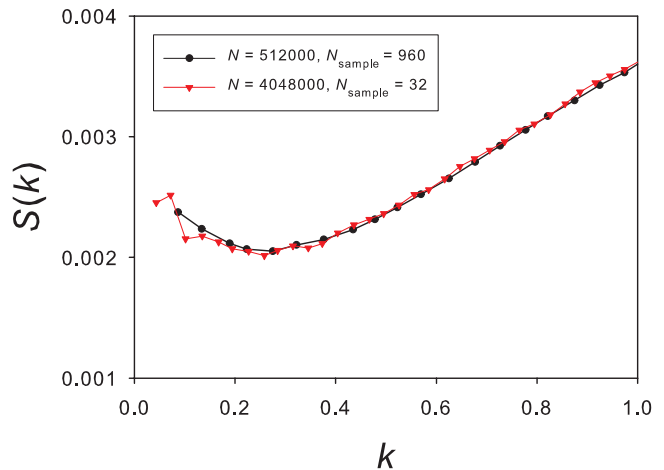,width=8.5cm}
\caption{The static structure factor at $\phi =$ 0.7 
for different system sizes, showing that the upturn near 
$k \approx 0.3$ found for $N=512000$ is not a finite-size effect.}
\label{fig:finitesize}
\end{center}
\end{figure}

Because we focus on the large-scale structure, 
it is very important to keep finite size effects under control. 
In Fig.~\ref{fig:finitesize} we show that the low wavevector 
crossover behavior observed near $k \approx 0.3$ for 
$N = 512000$ particles remains at the same position 
when using a larger system with $N = 4048000$ particles. 
The overall behavior of $S(k)$ including the upturn at $k \approx 0.3$
is unchanged, and thus it does not originate 
from any finite size artefact.
Note that the upturn at even lower wavectors seems even more pronounced 
in the larger system.
Similar data were obtained in two dimensions in Ref.~\cite{wu2015}.

\begin{figure}
\begin{center}
\psfig{file=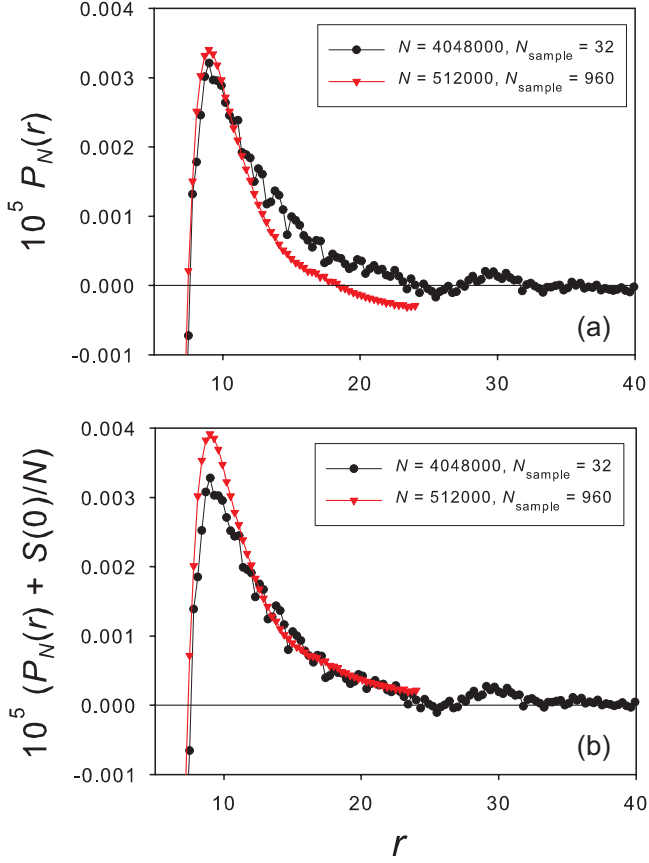,width=8.5cm}
\caption{\red{(a)} System size dependence of the coarse-grained 
density correlation function $P_N(r)$ at $\phi =$ 0.7. 
\red{(b)} System size dependence of the coarse-grained density 
correlation function 
with the finite size effect correction $P_N(r) + S(0)/N$ at $\phi =$ 0.7.
} 
\label{fig:finite}
\end{center}
\end{figure}

In Fig.~\ref{fig:finite} we show the system size dependence of the
pair correlation of the coarse-grained density field $P_N(r)$. 
We use the notation $P(r)$ only for the one in the infinite system size 
limit and 
$P_N(r)$ is used for the one measured with a finite 
$N$. 
In contrast to the case of $S(k)$, 
$P_N(r)$ exhibits a system size dependence.
The large-$r$ limit of $P_N(r)$ of the $N=512000$ system 
is clearly negative, 
while that of the $N=4048000$ system is closer to zero. 
Such behavior of the radial distribution function in thermal systems 
is a well-known finite size effect. 
We briefly review the method commonly-used to correct 
for this effect~\cite{salacuse1996}. 
In the grand canonical ensemble, 
the density-density correlation function for an open system 
$\ave{\rho(\vec{x}) \rho (\vec{x}+\vec{r})}$ 
can be written as 
\begin{eqnarray}
\ave{\rho(\vec{x}) \rho (\vec{x}+\vec{r})} = 
\sum_{N} p(N) \ave{\rho(\vec{x}) \rho (\vec{x}+\vec{r})}_N
\end{eqnarray}
where $\ave{\rho(\vec{x}) \rho (\vec{x}+\vec{r})}_N$ 
is the  density-density correlation function for a closed system with 
$N$ particles, 
and $p(N)$ is the probability that the open system contains $N$ 
particles~\cite{hansen}.  
When $N$ is sufficiently large, 
$p(N)$ becomes a function with a sharp peak at the most probable $N$, 
which we denote by $N^*$. 
Therefore, by expanding $\ave{\rho(\vec{x}) \rho (\vec{x}+\vec{r})}_N$ 
around $N = N^*$, and taking the summation over $N$, 
we obtain 
\begin{eqnarray}
G(r) = G_{N^*}(r) 
+ \frac{S(0)}{2N^*} \ppdif{\rho^2 G_{N^*}(r)}{\rho} + O((\frac{1}{N^*})^2),
\end{eqnarray}
where we introduce $G(r) = \rho^{-2} \ave{\rho(\vec{x}) \rho (\vec{x}+\vec{r})}$
and $G_N(r) = \rho^{-2} \ave{\rho(\vec{x}) \rho (\vec{x}+\vec{r})}_N$ 
as in Eq.~(\ref{eq:gofr}). 
The same derivation can be directly applied for 
the coarse-grained density-density correlation function 
$Q(r)$ as defined in Eq.~(\ref{eq:qofr}). 
The derivation may hold even in our athermal system 
if one assumes that $p(N)$ 
is still characterized by a sharp peak, as in thermal systems, which
is a reasonable hypothesis. 
Thus one can expect 
\begin{eqnarray}
Q(r) = Q_{N^*}(r) 
+ \frac{S(0)}{2N^*} \ppdif{\rho^2 Q_{N^*}(r)}{\rho} + O((\frac{1}{N^*})^2),
\end{eqnarray}
where $Q_N(r)$ is the value of $Q(r)$ for a closed system with 
$N$ particles.
Because we focus on the large-$r$ region, 
we can safely set $Q_{N^*}(r) = 1$ in the second derivative.
This leads to the finite size correction for $P(r)$: 
\begin{eqnarray}
P(r) \approx P_{N^*}(r) + \frac{S(0)}{N^*}, 
\end{eqnarray}
where $P_N(r) = Q_N(r) -1$. 
To check the reliability of this analysis, 
we plotted $P_{N}(r) + \frac{S(0)}{N}$ for $N=512000$ and $4048000$ 
in the bottom panel of Fig.~\ref{fig:finite}. 
The results obtained from different system sizes 
become nearly identical, thus the finite size effect is corrected. 
We use $P(r)$ obtained in this way in this work. 

\section{Number of samples}
\label{app:finitesample}
 
\begin{figure}
\begin{center}
\psfig{file=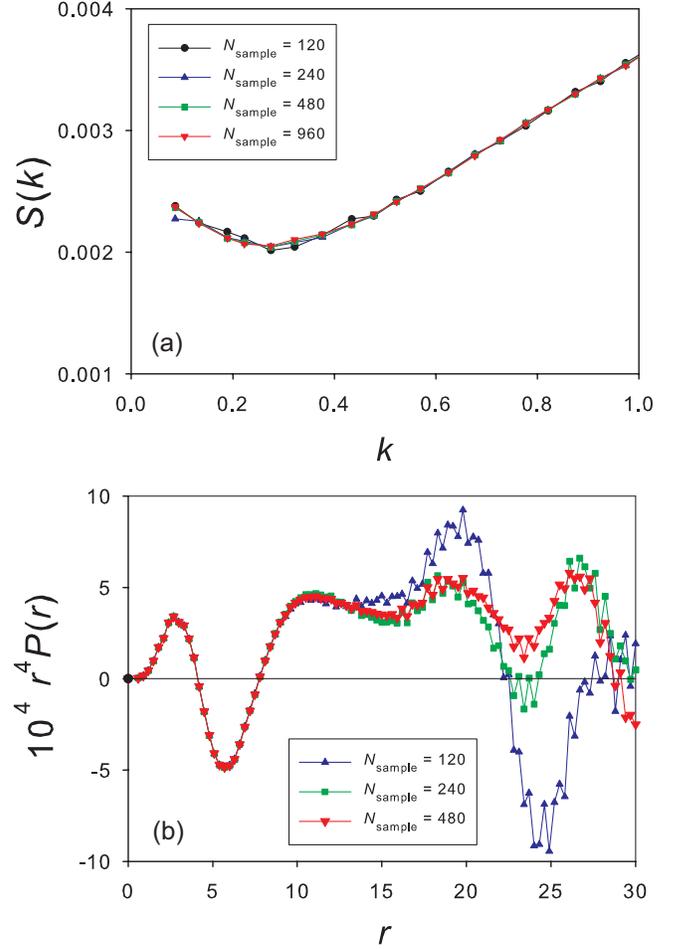,width=8.5cm}
\caption{Dependence on the number of \red{independent} samples of 
\red{(a)} $S(k)$ at $\varphi=0.7$ and \red{(b)} $r^4 P(r)$ at $\phi =0.66$.
These results show that the low-$k$ behavior 
of $S(k)$ and the large-$r$ behavior of $P(r)$ 
are statistically significant.}
\label{fig:sample}
\end{center}
\end{figure}

In Fig.~\ref{fig:sample} we show that our measured results for 
$S(k)$ and for $P(r)$ are obtained for a sufficiently large 
number of independent samples. 
The low-$k$ region of $S(k)$ and the large-$r$ region of $P(r)$ 
have relatively large sample-to-sample fluctuations, but 
they eventually converge very well when the number of samples 
is larger than 480. 
This test confirms the robustness of our data set with respect to the 
chosen number of independent samples.

\end{document}